\documentclass{MITcsail}
\usepackage{lipsum}
\usepackage{mathrsfs}
\usepackage{tikz}
\usepackage{epsfig}
\usepackage[ruled,vlined]{algorithm2e}
\usepackage{times}
\usepackage{graphicx,amsmath,amssymb,stfloats, url,amssymb}
\usepackage{multirow}
\usepackage{listings}
\usepackage{balance}
\usepackage{graphicx} 
\usepackage{float}
\usepackage{makecell}
\usepackage[mathscr]{eucal}
\usepackage{chngpage}
\usepackage{authblk} 

\newcommand{\stitle}[1]{\vspace{1ex}\noindent{\bf #1}}

\newcommand{\eat}[1]{}
\newcommand{\ie}{\emph{i.e.,}\xspace}
\newcommand{\eg}{\emph{e.g.,}\xspace}

\title{LLM4Hint: Leveraging Large Language Models for Hint Recommendation in Offline Query Optimization}

\author[1]{Suchen Liu\footnote{2301111946@pku.edu.cn}}
\author[1]{Jun Gao\footnote{gaojun@pku.edu.cn}}
\author[2]{Yinjun Han\footnote{han.yinjun@zte.com.cn}}
\author[2]{Yang Lin\footnote{lin.yang@zte.com.cn}}

\affil[1]{Key Laboratory of High Confidence Software Technologies, CS, Peking University, China}
\affil[2]{ZTE Corporation}

\begin{document}
\maketitle
\begin{abstract}
Query optimization is essential for efficient SQL query execution in DBMS, and remains attractive over time due to the growth of data volumes and advances in hardware. Existing traditional optimizers struggle with the cumbersome hand-tuning required for complex workloads, and the learning-based methods face limitations in ensuring generalization. With the great success of Large Language Model (LLM) across diverse downstream tasks, this paper explores how LLMs can be incorporated to enhance the generalization of learned optimizers. Though promising, such an incorporation still presents challenges, mainly including high model inference latency, and the substantial fine-tuning cost and suboptimal performance due to inherent discrepancy between the token sequences in LLM and structured SQL execution plans with rich numerical features.

In this paper, we focus on recurring queries in offline optimization to alleviate the issue of high inference latency, and propose \textbf{LLM4Hint} that leverages moderate-sized backbone LLMs to recommend query optimization hints. LLM4Hint achieves the goals through: (i) integrating a lightweight model to produce a soft prompt, which captures the data distribution in DBMS and the SQL predicates to provide sufficient optimization features while simultaneously reducing the context length fed to the LLM, (ii) devising a query rewriting strategy using a larger commercial LLM, so as to simplify SQL semantics for the backbone LLM and reduce fine-tuning costs, and (iii) introducing an explicit matching prompt to facilitate alignment between the LLM and the lightweight model, which can accelerate convergence of the combined model. Experiments show that LLM4Hint, by leveraging the LLM's stronger capability to understand the query statement, can outperform the state-of-the-art learned optimizers in terms of both effectiveness and generalization.

\end{abstract}

\section{Introduction}
\label{sec-intro}

The query optimizer is a critical yet challenging component within any database management system (DBMS), playing a central role in selecting the most efficient execution plan \citep{basicqueryop}. Given a query, the query optimizer enumerates candidate execution plans, and attempts to select the most efficient plan from the exponential number of candidates, among which the efficient one may be orders of magnitude faster than the poor one. With years of efforts in human-tuned rules for complex queries and diverse data distributions, traditional optimizers have achieved robust performance for different kinds of queries, but still need substantial work to capture complex relationships between data distribution and query patterns, and to adapt to enlarged data scale or emerging hardware architectures.

In response to these limitations, a range of learning-based approaches for query optimization have emerged~\citep{learnedsurvery23}, in which optimization policies are learned from feedback rather than hand-crafted. The preliminary step in the learning-based approaches is to capture the key features from data distribution and query execution plan~\citep{learnedcost2019, NeuroCard2020, QueryFormer2022}, using neural networks like TreeCNN~\citep{treecnn2016} and Transformer~\citep{vaswani2017attention}. Based on these key features, the learned optimizers construct efficient execution plans following the generation methods \citep{rtos2020, balsa, LOGER2023} or the selection methods \citep{bao2021,lero2023}, where the former methods attempt to build execution plans from scratch, and the plans considered in the latter methods come from the DBMS optimizer under various hints. These learned methods have achieved better performance after sufficient training. Despite these advances, learned optimizers still suffer from weaknesses in terms of generalization, \ie incoming queries are different from the training ones. Moreover, the existing learning methods have limited capability in understanding the query statement, such as the specific conditions on the query predicates~\citep{QueryFormer2022}.


Recent years have witnessed the remarkable success of LLMs \citep{kenton2019bert, gpt, gpt2, gpt3, gpt4}. Trained on massive corpora, LLMs with a large number of parameters have demonstrated impressive capabilities and generalization in various NLP downstream tasks and application domains. It is reasonable to expect that the knowledge encoded by LLMs can enhance existing methods for better performance and generalization in query optimization. In fact, LLMs have shown promising results in other database tasks, especially text-rich tasks, such as table understanding~\citep{TableMeetsLLM2024,liu2021tapex,CoT2022}, text-to-SQL~\citep{zhang2024benchmarkingtext2sql,Bird,pourreza2024din,li2023resdsql,gao2023text,wang2023mac,li2023graphix}, index tuning~\citep{DB-GPT}, and database diagnostics~\citep{D-Bot, LLMDB}.

Although LLMs have the potential to overcome the limitations of existing learned optimizers, we cannot directly apply LLMs to query optimization due to the following challenges. First, LLM-based methods obviously incur \textbf{higher inference latency} compared to traditional or even existing learning-based methods. LLM typically contains numerous parameters with multiple layers, resulting in significant inference time. When the underlying data is relatively small, the time consumed by the online query optimization will exceed the time saved during query execution, resulting in increased overall latency.

Second, the mismatch between token sequences in LLMs and tree-structured SQL execution plans generated by the optimizer results in \textbf{substantial fine-tuning cost} and \textbf{suboptimal performance} even after extensive fine-tuning. Most existing LLMs are primarily trained on input token sequences to generate output tokens by considering all previous tokens~\citep{gpt}. In such sequences, tokens have no explicit relationships with each other, while query execution plans are explicitly structured as trees. Additionally, LLMs treat numerical data as tokens without fully capturing their continuous characteristics~\citep{NumberLLM}, resulting in difficulty in performing tasks that require comparisons of numerical data. Moreover, LLMs need sufficient statistics, like data distribution, to make correct decisions during plan generation~\citep{basicqueryop}. These distributions vary across different datasets and have to be fed into LLM at runtime, leading to a long input context. These factors necessitate substantial fine-tuning when employing LLMs for query optimization, as execution plans represent a relatively unfamiliar data modality for LLMs.

Regarding the first challenge, this paper focuses on a more practical application scenario: optimizing repeated SQL execution in the offline setting~\citep{offlineop}, which is insensitive to the time consumed in query optimization. In fact, many enterprises routinely run pre-defined jobs containing SQLs on a daily basis. For example, studies~\citep{RepeatSQL2024} show that 80\% of queries in Amazon Redshift's workloads are repeated. When notified that certain queries take excessive time, managers can perform offline optimization, \eg by generating appropriate query hints~\citep{pghint}. These hints, obtained once, will improve the performance of these queries in the following repeated executions. We leave the online learned query optimizer for future work, once LLM inference becomes significantly more efficient.

Regarding the second one, we observe that existing studies of enhanced LLMs to support tasks like table understanding~\citep{tableGPT2, MTableACL2024, DocLayLLMCVPR2025} and time series forecasting~\citep{temporalLLM2023,TimeLLM,challu2023nhits} face similar difficulties. To enrich LLMs with knowledge in specific domains effectively, these methods introduce lightweight models to handle the specific data, and align the embedding space of the lightweight models with the embedding space in LLM. This paper follows the similar idea to combine a lightweight model into LLM, which then can handle the tree-structured query execution plan with rich numerical features, and reduce the input context length for the LLM. However, expensive fine-tuning is still needed for the combined offline query optimizer, as the query plans, with nested tree structure and more heterogeneous features, are much more complex than tabular and time series data.

Taking into account the aforementioned factors, we propose \textbf{LLM4Hint}, a framework that leverages LLM into query evaluation to improve the efficiency and generalization of learned query optimizers. Given a query, LLM4Hint enumerates candidate execution plans guided by different database hints~\citep{pghint}, introduces a specific lightweight model to capture the tree-structured query execution plan, and relies on the LLM as the backbone to understand the textual feature of SQL and outputs of the lightweight model, so as to choose the efficient plan (or corresponding hint) by pairwise comparison. The key contributions of this paper are summarized as follows:

\begin{itemize}
\item LLM4Hint introduces a lightweight model to encode rich numerical features in data distribution and query execution plans into space-concise embeddings, and uses these embeddings as the soft prompt to provide necessary statistics to the LLM. After fine-tuning, the soft prompt is aligned with the embedding space of the LLM, which enables the LLM to produce more efficient results.

\item We propose various optimization strategies to improve performance while controlling fine-tuning costs. First, we develop a SQL rewriting strategy to convert SQL into natural language by a larger LLM, so as to facilitate the moderate-sized LLM to understand the SQL text easily. Second, we introduce the matching text into the LLM context to explicitly express the relationships between the soft prompt generated by the lightweight model and the tables in the SQL statement, which lowers the alignment cost and enhances generalization ability.

\item Extensive experiments conducted on multiple datasets demonstrate that LLM4Hint outperforms existing learned query optimization methods in terms of both efficiency and generalization.

\end{itemize}

The remainder of this paper is organized as follows. We review the related work in Section~\ref{sec-related}. Then, we describe LLM4Hint in Section~\ref{sec-method}. Section~\ref{sec-exp} reports the experimental results. We conclude the paper in Section~\ref{sec-conclusion}.

\section{Related Works}
\label{sec-related}

We review related works including the representation of the execution plan, the learned query optimizer, the advances of LLMs and the adaption of LLMs into different tasks in DBMS, and the enhanced LLM for table understanding and time series analysis.

\subsection{Representation of Execution Plan}

The accurate representation of a query execution plan is a basic step for query optimization. Early efforts have utilized operator-specific neural units for various operators, such as scans and joins, and have integrated these units based on the execution plan's architecture~\citep{singleop2019}. Hierarchical models like Tree-LSTM~\citep{treelstm} and Tree-CNN~\citep{treecnn2016} accumulate information from the leaf nodes along the paths of the tree to form the representation of the plan. QueryFormer~\citep{QueryFormer2022} expands upon multi-layer transformers for the representation of query plans, with each node annotating the related database statistics and features of predicates, to enable the precise cost estimation with an increased number of transformer parameters.

\subsection{Learned Optimizer}

Learned database optimizers have become an active area of research in recent years~\citep{learnedsurvery23}, broadly categorized into generation and selection approaches. Most of the generation methods formulate the plan generation as a sequence decision problem, and learn how to determine the next table to be joined or the physical join operator. Among them, RTOS~\citep{rtos2020} is based on a value-based reinforcement learning framework, employs Tree-LSTM to model intermediate evaluation subtrees, and investigates the estimated database cost to guide the model pre-training. Following the similar framework of RTOS, LOGER~\citep{LOGER2023} further reorganizes the search space by disabling certain operators, and devises a beam search method under a combined loss function that considers both the join order and the physical operator selection. Balsa~\citep{balsa} implements a policy-based reinforcement learning approach, and trains on a simulator before fine-tuning with real-world execution data. These learned methods can generate more efficient plans than those from DBMS after full training, but face performance fluctuation when an incoming query is unlike the training queries.

The learned selection methods, represented by Bao~\citep{bao2021} and Lero~\citep{lero2023}, first compose different hints to notify the DBMS optimizer to produce diverse plans, and then select one efficient plan using a data distribution aware value model. The differences among the existing methods lie in the kinds of hints (\eg disabling or forcing specific implementations~\citep{bao2021}, correcting estimated row numbers~\citep{lero2023}), and the ways to select the plans (\eg absolute time~\citep{bao2021}, relative comparison~\citep{lero2023}). Fastgres~\citep{fastgres2023} learns to select hints directly on the cluster of similar queries using the traditional machine learning method \ie GDBT. Compared to generation methods, the selection methods achieve better generalization to unseen queries, as the plans generated from the DBMS optimizer are generally of acceptable quality even under different hints. LLM4Hint follows Lero's setting ~\citep{lero2023} to select the hints based on the relative cost comparison between different plans, and expects that the understanding capabilities of LLM can boost the effectiveness of selected hints and improve the model's generalization.

\subsection{Large Language Models}

The transformer architecture~\citep{vaswani2017attention} has enabled models like BERT~\citep{bert} and the GPT series~\citep{gpt2,gpt3,gpt4} to excel in NLP tasks such as translation, summarization, and question answering. Recent models like GPT-4~\citep{gpt4} and Llama~\citep{llama2} further enhance LLMs' abilities in encoding open-world knowledge and multi-step reasoning. After receiving user input and prompts in context, LLMs generate output by identifying key factors and leveraging internal knowledge. LLMs with longer context windows generally exhibit improved capabilities, but with higher training costs.

\eat{There are different methods to adapt the LLM for specific tasks. Fine-tuning is a direct one, among which the full fine-tuning strategy updates all parameters, and the partial fine-tuning strategy focuses on the parameters in specific layers. As the parameters in terms of weight matrix $W$ in each layer are still large, LoRA~\citep{lora} is proposed to further reduce the fine-tuning cost. Specifically, LoRA keeps $W$ in LLM unchanged, and introduces two small learned matrices $A$ and $B$ with rank $r$, whose product serves as the modification part of $W$ in the specific task. Such an approach significantly reduces the number of parameters that need to be trained (\ie only including the parameters in $A$ and $B$), which makes the training process faster and more manageable.}

There are different methods to adapt LLMs for specific tasks. Fine-tuning is a straightforward approach, among which the full fine-tuning strategy updates all parameters, and the partial fine-tuning strategy focuses on the parameters in specific layers. As the parameters in terms of weight matrix $W$ in each layer are still large, LoRA~\citep{lora} is proposed to further reduce the fine-tuning cost. Specifically, LoRA keeps $W$ in LLM unchanged, and introduces two low-rank matrices $A$ and $B$ such that their product approximates the change in $W$ during fine-tuning. The rank of matrices $A$ and $B$ is determined by a hyperparameter $r$, which balances the model complexity and the ability to capture task-specific information. Such an approach significantly reduces the number of parameters that need to be trained (\ie only including the parameters in $A$ and $B$), which makes the training process faster and more manageable.

\eat{adjacent to the pre-trained weight matrix $W$, without altering the parameters of $W$. This approach significantly reduces the number of parameters that need to be trained, making the training process faster and more manageable. The low-rank matrices A and B are initialized such that their product approximates the change in the weight matrix W during fine-tuning. The rank $r$, which is a hyperparameter, determines the size of these matrices and thus the trade-off between model complexity and the ability to capture task-specific information. LoRA has been shown to reduce the parameter count significantly while maintaining or even surpassing the performance of full parameter fine-tuning. LoRA is particularly useful when frequent task switching is required, as LoRA allows for quick and memory-efficient updates to the model's behavior.}

Prompt tuning methods are alternatives to model fine-tuning. The prompt, along with other input, is used to invoke the internal capability of LLMs without updating their parameters. Among them, hard prompt means discrete text tokens meaningful to humans, which can be designed by an expert, or mined to improve performance~\citep{autoprompt}. Since it is non-trivial to locate the meaningful conditioning discrete text as hard prompt, Prompt Tuning \citep{prompttunning} introduces $k$ learned embeddings as soft prompt, and feeds the soft prompt along with other text prompts to LLM. The soft prompt, trained end-to-end, can achieve better performance without human involvement. LLM4Hint also follows the soft prompt approach, which incorporates the embeddings representing the given query evaluation plan into the LLM context.

The database field also embraces the advancement of LLMs. LLMs have been explored~\citep{LLMDB} for providing commonsense knowledge, reducing the human burden and improving performance in various tasks, like table understanding~\citep{TableMeetsLLM2024,liu2021tapex,CoT2022,iida2021tabbie,gong2020tablegpt,yang2022tableformer,nassar2022tableformer}, query rewriting~\citep{DB-GPT}, index tuning~\citep{DB-GPT}, and database diagnostics~\citep{D-Bot}, and text-to-SQL tasks~\citep{zhang2024benchmarkingtext2sql,Bird,pourreza2024din,li2023resdsql,gao2023text,wang2023mac,li2023graphix}. These tasks typically involve rich and flexible descriptive text, making them well-suited for LLMs. As we have mentioned in the introduction, the query optimization task, with its distinct characteristics like explicit tree-structured plans with rich numerical features, poses more challenges for the adaptation of LLMs compared to other tasks.

\subsection{Enhanced LLM for Time Series and Tabular Data}

LLMs have been studied and explored in time series analysis, where the time series data exhibits significant differences from the training corpora of LLMs. In addition to methods that treat series data directly as text prompts for LLMs, some methods attempt to build embeddings for time series and align these embeddings with the LLM space. Among them, GPT4TS \citep{gpt4ts} uses patch embeddings~\citep{patch} for input raw data, and then fine-tunes the specific layers (\eg embedding layers) while retains other layers (\eg self-attention layers) in GPT-2 to support various analysis tasks. Time-LLM~\citep{TimeLLM} annotates different time series patches using a vocabulary subset (\eg \textit{up, long}) related to temporal features, which serves as easily understood prompts for LLMs. Time-LLM also provides other text prompts, including task-specific instructions and required statistical data. 

The enhancement of LLM to understand tabular data also receives great attention. With the explicit row/column structure, tabular data differs from the plain text that is fed to the LLMs. TableGPT2~\citep{tableGPT2} views the table data as a new modality, and incorporates a table encoder capturing schema-level and cell-level information during fine-tuning LLM. The works on multi-modal table understanding~\citep{MTableACL2024} and DocLayLLM~\citep{DocLayLLMCVPR2025} handle the table understanding from images based on the multi-modal LLM, in which encoders capture text, layout, visual and other key features, and align the embeddings with those in the backbone LLM during fine-tuning. These methods achieve competitive results in enhancing LLM to support their respective data modalities.

Inspired by these works, we integrate a lightweight model to encode the query execution plan to handle numerical features and the tree-structured plan. Considering the unique structures and semantic requirements of SQL queries and execution plans, we also explore the method to rewrite SQL and explicit matching prompt to make LLM easily understand the task and needed features. 

\section{LLM4Hint}
\label{sec-method}

In this section, we first discuss different design choices in applying LLMs to query optimization and then give the overall framework of LLM4Hint. We further detail three core components of LLM4Hint, including a larger LLM enhanced SQL text, a lightweight model to encode features of the query execution plan, and an explicit matching prompt to facilitate alignment of the LLM and lightweight model. We finally show the training loss function, and present the complete training and inference of LLM4Hint.

\subsection{Design Choices}

\eat{We relist the challenges faced by LLMs, including (i) the mismatch between the token sequence and the tree-structured SQL execution plan, (ii) the long context needed for the optimization statistics and expensive model fine-tuning, and (iii) limited understanding of the numerical data. }

Before delving into the detailed components of LLM4Hint, we first explain our key design choices to achieve an effective LLM-based offline query optimizer with relatively low fine-tuning cost.

\stitle{A Single LLM or a Combined LLM-Lightweight Model.} Facing these challenges, one option is to fine-tune a large LLM to produce the query execution plan. As mentioned before, the necessary optimization statistics have to be fed into the LLM at runtime, indicating that the LLM needs to support long context and the subsequent fine-tuning cost is high. Although promising, such an approach requires too many computing resources, like distributed GPU training. Instead, we choose a combined model ~\citep{gpt4ts, TimeLLM} that includes both an LLM and a lightweight model, in which the former is used to understand SQL statements, and the latter, similar to the existing query execution plan encoder~\citep{QueryFormer2022}, is used to represent the features of the query execution plan, which takes the tree form with numerical features. The lightweight model will yield a sequence of embeddings, and append these embeddings as the soft token ~\citep{prompttunning} to the prompt. As mentioned above, the combined model can partially address the challenges with limited computing resources, \eg one GPU card.


\stitle{Strategies in Handling Cross Model Relationship.} The introduction of a lightweight model raises another issue, \ie the embedding generated by the lightweight model belongs to a space different from that of the LLM. Cross-alignment between two spaces~\citep{gpt4ts, TimeLLM}, although feasible, requires many training instances. We take two strategies to handle this issue. Instead of producing a single plan embedding~\citep{QueryFormer2022}, we provide a sequence of embeddings (\eg covering all tables and operators) to enrich the context and facilitate fine-tuning easier with more features. In addition, we express the explicit relationships between the SQL statement and the embeddings produced by the lightweight model as part of the text prompt, reducing the fine-tuning cost and enhancing generalization ability.

\stitle{Choices of the Backbone LLM.} The full understanding of SQL optimization requires powerful LLMs~\citep{gpt4}. As we have introduced the lightweight model to capture the key features of the execution plan, the context required for the prompt can be greatly reduced, which offers the chance of using a moderate-sized LLM in fine-tuning. Additionally, larger LLMs cannot gain advantage without spending. In addition to their expensive fine-tuning and inference, larger LLMs incur an access fee. For example, GPT-4~\citep{gpt4} costs $\$$0.03 per 1K input tokens and $\$$0.06 per 1K output tokens for its 8K context version currently. Considering two factors, we choose a moderate-sized LLM as the backbone of the LLM4Hint. Specifically, similar to the time series analysis method~\citep{gpt4ts}, we mainly use a GPT-2~\citep{gpt2} version with 117M parameters as the backbone LLM of LLM4Hint to balance effectiveness and training cost. 

\eat{Additionally, we devise an SQL rewriting strategy using a larger LLM to enable the moderate-sized LLM to better understand the SQL statement.}

\eat{We also perform studies on other similar-sized models, including Qwen, and Llama, in our experimental studies.}

\stitle{Choices of Query Optimization Tasks.} Different existing learned optimizers handle different tasks for query optimization, such as selecting of the next table to be joined in the execution plan construction~\citep{rtos2020,LOGER2023}, and selecting specific hints (\eg enable hash join, disable sort-merge join) to generate the efficient execution plans~\citep{bao2021, lero2023}. This paper focuses on the latter issue in offline optimization setting, and leaves other tasks in the future work. Specifically, we choose the relative cost comparison~\citep{lero2023} instead of the absolute time estimation~\citep{bao2021} in selecting proper hints, as the relative cost is the key factor in the query optimization~\citep{lero2023}.

\eat{The selection methods search among a fixed space based on the entire plan or query, while the generation methods have to encode intermediate subplans with the dynamic output.}

\subsection{Architecture}
\begin{figure*}[!t]
    \centering
    \includegraphics[width=\linewidth]{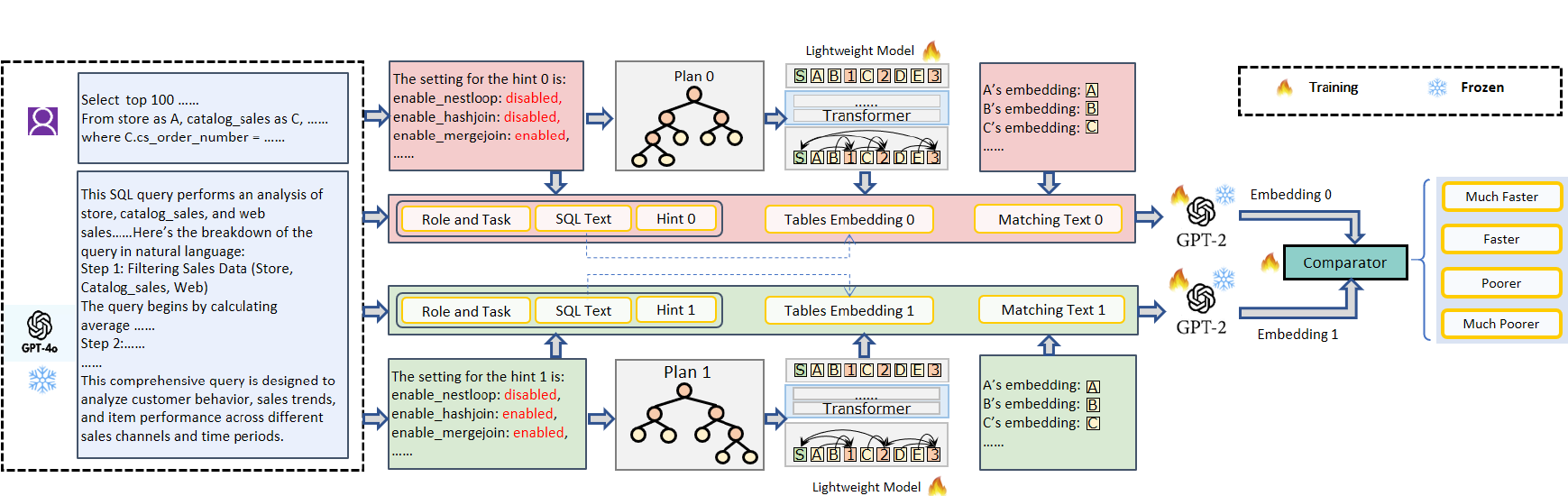}
    \caption{Framework of LLM4Hint}
    \label{fig:framework}
\end{figure*}
Figure~\ref{fig:framework} shows the LLM4Hint framework. Given a query and an optimization hint, LLM4Hint invokes DBMS to generate a query execution plan, and takes a lightweight model $\mathcal{M}_p$ to convert the plan into a sequence of embeddings. The complete prompt for the LLM is then composed of the hard prompt and the soft prompt, where the former includes SQL text and other text prompts, and the latter is the generated embeddings of the lightweight model. A moderate-sized LLM $\mathcal{M}_b$ summarizes the context into an embedding, which serves as the final embedding of the plan. Similarly, the embedding for a plan under another hint will be produced. As in Lero~\citep{lero2023}, two plan embeddings are fed into a comparator to determine the relative advantage between the two plans, from which the more efficient one (or the corresponding hint) can be selected finally. The symbols commonly used throughout this paper are listed in Table \ref{tab:symbols}.

The key part of LLM4Hint is the composition of different prompts for $\mathcal{M}_b$, including the enhanced SQL text, the soft prompt for the execution plan generated from the lightweight model $\mathcal{M}_p$, and the matching relationship between $\mathcal{M}_b$ and $\mathcal{M}_p$.

\begin{itemize}

\item \textbf{Enhanced SQL Text:} The SQL statement part includes the part to specify the role and the task, the plain SQL in the text form, and the hint to guide the execution plan generation. As the moderate-sized LLM has limited capability in understanding the SQL query, we optionally leverage a more advanced LLM to convert the SQL query to more friendly natural language sentences, which can better suit $\mathcal{M}_b$.

\item \textbf{Soft Prompt for Execution Plans:} Given a query and a hint, we use the statement {\sl Explain} to obtain the corresponding query evaluation plan. We borrow the state-of-the-art QueryFormer~\citep{QueryFormer2022} as the lightweight model to convert the tree-structured plan into a sequence of embeddings, which empowers the backbone LLM $\mathcal{M}_b$ to capture the needed statistics more precisely. Take Figure~\ref{fig:framework} as an example. The soft prompt contains the embeddings in the last layer in QueryFormer, like embeddings for the super node, tables aliases and operators.

\item \textbf{Matching Relationships between Two Models:} The matching part attempts to express the relationships between the SQL text and the soft prompt explicitly, which can lower the alignment costs between the two models. Take Figure~\ref{fig:framework} as an example. The matching text directly presents the correlative relationships between the table embedding and the corresponding table in the SQL expression.

\end{itemize}

LLM4Hint training requires the update of parameters in both the lightweight model $\mathcal{M}_b$ and the comparator $\mathcal{M}_c$, in which the parameter optimization in $\mathcal{M}_b$ not only captures the plan features but also aligns its embedding space with the LLM space. As for the moderate-sized LLM $\mathcal{M}_b$, $\mathcal{M}_b$ can be frozen or fine-tuned using LoRA~\citep{lora} in the top transformers of the LLMs. We have performed different tests in the experimental study.

\begin{table}[!t]
\caption{Frequently Used Symbols}
\centering
\begin{tabular}{|c|p{.35\textwidth}|}
\hline
Symbols &  Meaning \\
\hline
$q$ & The query\\
$h_0$, $h_1$ & Two hint sets $h_0$ and $h_1$\\
$p_q^{h0}$, $p_q^{h1}$ & The plan for $q$ constructed with the guide of $h_0$ and $h_1$\\
$e_q^{h0}$, $e_q^{h1}$ & The representation embedding for $p_q^{h0}$ and $p_q^{h1}$\\
$\mathcal{M}_b$ & The moderate-sized backbone LLM \\
$\mathcal{M}_l$ & The larger LLM for the SQL-to-NL\\
$\mathcal{M}_p$ & The lightweight model to encode the features of execution plan\\
$\mathcal{M}_c$ & The plan-wise comparator model\\
\hline
\end{tabular}
\label{tab:symbols}
\end{table}

\subsection{Larger LLM Enhanced SQL Text Statement}

The SQL text statement includes three parts: the role specification part, the SQL text part, and the hint part.

\stitle{Role Specification Part.} Role specification part serves as a basic guide used in the specific task. In LLM4Hint, such a prompt takes the form as: {\it You are a helpful assistant specializing in query optimization. You need to capture the key feature of the SQL execution plan.}

\stitle{SQL Text Part.} We incorporate the SQL text form into the prompt for two primary reasons. First, LLMs excel at the task of understanding natural language.
To harness the full capabilities of LLMs, the SQL text form should better be included into the prompt. Second, while the lightweight model, QueryFormer~\citep{QueryFormer2022}, can capture key features necessary for cost estimation, the lightweight model still lacks the support for complex expressions, especially the cardinality estimation on string filtering operations like "\textit{LIKE}" and "\textit{IN}". The absence of these features could potentially affect the accuracy of cost estimation. Therefore, we incorporate the SQL text component into the prompt as a supplementary source to address this limitation.

Note that although the original SQL query takes the plain text form that can be directly added to the prompt, the SQL queries, with the rigid syntax requirement, are still not friendly to the moderate-sized LLM, which is mostly pre-trained on the natural language in the training corpora.

LLM4Hint then introduces a query rewriting strategy aided by a larger LLM to convert the SQL statement into natural language text, which is better suited to LLM's capabilities. Given an SQL query $q$, we use a larger LLM $\mathcal{M}_l$, \eg GPT-4~\citep{gpt4}, and invoke a rewriting task with the following prompt: {\it
You are an SQL explanation assistant. I will provide an SQL query, and you should convert it into a brief natural language description, summarizing its purpose and steps. Highlight the goal of the query, the involved tables, any filtering conditions, and any aggregation, sorting, or other operations used. SQL: \{SQL query\}.} The conversion results, as we can see from Figure \ref{fig:framework}, are detailed steps for the given queries. These rewritten queries eliminate the specific SQL syntax while preserving the core intent in a more user-friendly expression, enhancing the comprehension of the moderate-sized LLM $\mathcal{M}_b$. We also note that $\mathcal{M}_l$ is invoked once for each query, which does not result in substantial access fees.

\stitle{Hint Text Part.} We include the hint text in the prompt with the expectation that the LLM can effectively model the relationships between the hint, the SQL statement, and the corresponding execution plan. Specifically, we select the first 16 hint combinations out of a total of 48, as identified in the study by Bao~\citep{bao2021}, where each hint represents a binary decision on whether a specific query execution strategy (such as the join method or scan type) is enabled or disabled. An example of the hint text part is shown in Figure~\ref{fig:framework}. Since query execution plans are influenced by the chosen hints, the fine-tuned LLM, denoted $\mathcal{M}_b$, is likely to learn implicit patterns, \eg for some query patterns (in terms of both SQL text and plan features), the specific hint may produce better (or poor) performance, which consequently enhances the model generalization.
\eat{we select the top 16 effective hint combinations out of a total of 48,}

\subsection{Lightweight Model for Tree-Structured Execution Plan}

The query execution plan takes a tree form, where the leaf nodes represent tables with their access methods (\eg table or index scans), the internal nodes correspond to operators (\eg joins, scans, and filters), and the edges define the order of operations, including the join sequence. As each operator produces an intermediate table, all nodes in the query plan tree are annotated with rich numerical features, including the cardinality of the nodes and the estimated running time cost. The estimated time of the entire plan is then computed using all nodes in the tree. Note that the estimated cardinality and cost from DBMS are not always accurate. Otherwise, we can simply use the estimated cost to select the most efficient plan.

QueryFormer~\citep{QueryFormer2022} estimates the execution cost with multiple transformer layers to process tree-structured plans. All nodes, including leaf and intermediate nodes, are tokens for the transformer. The nodes' embeddings are initialized via a linear layer on the collected features, including histograms in the database, the distribution of sampled data, and the estimated selectivity for the SQL predicates. The propagation of information along the transformer layers is not arbitrary, but restricted along the query execution tree, which is expressed by the structure mask matrix during the attention computation. A special node, the super node, is introduced into the transformer to collect all features for the entire execution plan.

Two extensions are introduced to integrate QueryFormer into LLM4Hint. First, the original work relies on only one embedding of the super node to make prediction, while in our context (about 1024 tokens in GPT-2), one embedding (or one token) provides too limited information. In order to enrich more detailed information, we choose all embeddings in the last layer of the transformer and use these embeddings as soft prompt. Second, inspired by multi-modal task approaches, LLM4Hint introduces an additional projection layer, implemented as a multi-layer perception (MLP), to map all outputs of the lightweight model to the LLM's vector space, so as to lower the cost in the alignment.

We illustrate the computation of the soft prompt in Figure~\ref{fig:match}. Suppose that the query statement contains $n$ table aliases. In the transformer, there will be at least $2n$ tokens: $n$ tokens for the leaf nodes, at least $n-1$ tokens for the intermediate nodes representing internal operations, and 1 token for the super table. After applying the tree-structured aware transformation layers, the embeddings retain key features of both the table nodes and the intermediate nodes, capturing the join relationships and the execution flow, respectively. 

\subsection{Matching Relationship Between Two Models}

The prompt now includes the enhanced SQL text and the soft prompt generated by the lightweight model $\mathcal{M}_p$. We believe that the embedding space between the backbone LLM $\mathcal{M}_b$ and $\mathcal{M}_p$ can finally be aligned after sufficient training, so $\mathcal{M}_b$ can be aware of the data distribution and query features. However, such a method incurs too much training cost and lacks generalization. For example, when different aliases are given to the same table, the learned model may fail to reuse the previously learned patterns to make a proper estimation.

\begin{figure}[!t]
    \centering
    \includegraphics[width=0.8\linewidth]{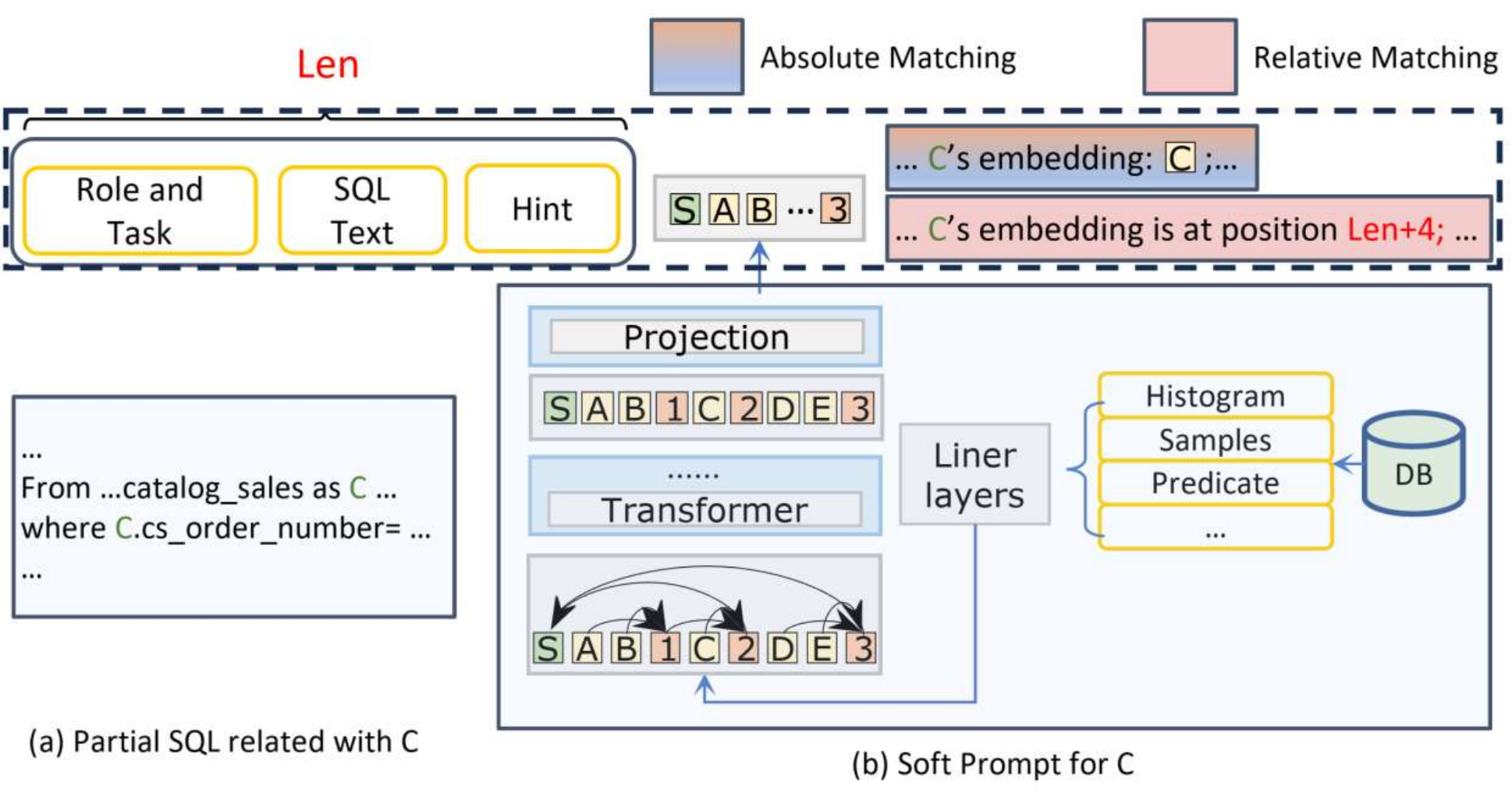}
    \caption{Illustration of Soft Prompt and Matching Text for Table $C$}
    \label{fig:match}
\end{figure}


In fact, there is no need to learn alignment relationships from scratch, as the table tokens fed into the transformer are extracted directly from the SQL query by ourselves, and the relationships are already known in advance. We can provide the explicit matching text for table nodes to accelerate alignment between the two models, enhancing generalization, such as the ability to handle unseen aliases. Besides, we do not generate matching text for the intermediate nodes, as intermediate nodes do not appear explicitly in the SQL text.

We propose two methods to express these relationships explicitly in the prompts. Take the table $C$ as an example. The relative relationship can be expressed as: {\it Table $C$'s embedding is at the position $p_C$}, where $p_C$ is the index of the soft prompt for $C$ in all tokens in the prompt. We demonstrate the computation of $p_C$ for table $C$'s embedding in Figure \ref{fig:match}. The process involves: (i) calculating the token length ($len$) in the enhanced SQL text segment, and (ii) identifying the position index ($idx$) of table $C$ in the query execution plan for transformer processing. The final position $p_C$ is the sum of $len$ and $idx$, which will be provided explicitly in the prompt. The relative position remains unchanged during the training, but is incompatible with the similarity computation in the transformers within the LLM.

\eat{We illustrate the computation of $p_C$ for the embedding of table $C$ in Figure \ref{fig:match}, in which we first determine the length of the tokens, denoted as $len$, in the initial enhanced SQL text part, and then locate the index of table $C$, denoted as $idx$,  in the query execution plan for the transformer.}

Another method is to express the absolute relationship as: {\it Table $C$'s embedding is $e_c$}, where $e_c$ duplicates the embedding collected at the top transformer layer for table $C$. Though $e_c$ may be updated during the training, LLM can easily establish the relationships between the table name and its soft prompt in the similarity computation in transformer, as the duplicated table name and duplicated embedding have appeared in the extended SQL part and the soft prompt, respectively. From experimental results, the absolute relationship can achieve better results than the relative relationship. Thus, LLM4Hint chooses the absolute relationship by default.

\subsection{Relative Advantage Aware Comparator}

LLM4Hint adopts a pair-wise training loss, so as to effectively differentiate the performance of various query execution plans, especially when faced with limited training data. As shown in Figure \ref{fig:framework}, the comparator is designed to assess the relative quality of two execution plans $p_q^{h0}$ and $p_q^{h1}$, based on their representation embeddings $e_q^{h0}$ and $e_q^{h1}$ collected at the top transformer layer in the LLM.

\eat{However, instead of a two-class comparison, we extend this to four categories.}

More fine-grained classes are necessary for comparing query execution times. Given a query, different execution plans can vary in performance by orders of magnitude. The 2-class classifier only shows which execution plan is faster or poorer, but doesn't indicate the extent of the difference. Therefore, in addition to the labels "faster" and "poorer", we introduce two additional labels: "much faster" and "much poorer" to capture the significant differences between the two execution plans.

Specifically, we concatenate two embeddings $e_q^{h0}$ and $e_q^{h1}$ into $e_c(0,1)\leftarrow e_q^{h0} || e_q^{h1}$. The comparator uses a combination of a router and two MLP classifiers to process the concatenated embedding. The router directs the concatenated embedding to the corresponding classifier, which then predicts the relative performance between the two execution plans. The final output $prob$ consists of four probability scores, representing the likelihood of one plan outperforming the another across four distinct categories:

\begin{equation}
\label{eq:comparator_prob_train}
    prob = \begin{cases}
        W_0 \cdot (e_c(0,1)) + b_0, f_{\textbf{Router}}(e_c(0,1)) =0\\
        W_1 \cdot (e_c(0,1)) + b_1, f_{\textbf{Router}}(e_c(0,1)) =1
        \end{cases}
\end{equation}

The ground-truth 4-class labels can be located with the actual execution time $t(p_q^{h0})$ and $t(p_q^{h1})$ between two plans $p_q^{h0}$ and $p_q^{h1}$ in the Equation~\ref{eq:label_truth}.
We define two thresholds. The first is the ratio $r_0 = t(p_q^{h0}) / t(p_q^{h1})$, and the second is the difference $r_1 = t(p_q^{h0}) - t(p_q^{h1})$.


\begin{equation}
\label{eq:label_truth}
y = 
\begin{cases} 
0\ (\text{much faster}), & \text{if } r_0 \geq 3/2 \ \mathbf{or} \ r_1 \geq 1000 \\ 
1\ (\text{faster}),       & \text{if } r_0 \in \left[1, 3/2\right) \ \mathbf{and} \ r_1 \in [0, 1000) \\ 
2\ (\text{poorer}),      & \text{if } r_0 \in \left[2/3, 1\right) \ \mathbf{and} \ r_1 \in [-1000, 0) \\ 
3\ (\text{much poorer}), & \text{if } r_0 < 2/3 \ \mathbf{or} \ r_1 < -1000 
\end{cases}
\end{equation}

We take the cross-entropy as the loss function, comparing the predicted probabilities and the ground-truth labels. Labels indicating a sharper performance difference are assigned greater importance. The weighted loss function can be described in Equation \ref{eq:loss}, where $l_n$ represents the loss for the $n$-th sample, $y_n$ is the true class label for the $n$-th sample, $prob_{n,k}$ is the predicted probability of label $k$ for the $n$-th sample, $w_{y_n}$ is the weight for the true class of the $n$-th sample, and $N$ is the total number of samples. 

\begin{eqnarray}
\label{eq:loss}
\mathcal{L} &=& \frac{1}{N} \sum_{n=1}^{N} w_{y_n} \cdot \ell_i \\
&=& -\frac{1}{N} \sum_{n=1}^{N} w_{y_n}\log{\frac{\exp(prob_{n,y_n})}{\sum_{k=0}^{C}{\exp(prob_{n,k})}}}
\label{eq:loss1}
\end{eqnarray}

\subsection{Overall Training and Inference Algorithm}

Finally, we present the overall training and inference algorithms. Algorithm~\ref{alg:training_LLM4Hint} details the training process of LLM4Hint, in which the query set $Q$ represent the training query set, hint set $H$ is chosen from the hint combinations identified in Bao~\citep{bao2021}, the backbone LLM $\mathcal{M}_b$ is the LLM for producing the plan embedding, a larger LLM $\mathcal{M}_l$ is for the conversion of the query into more natural text, a lightweight model $\mathcal{M}_p$ is for the modeling of the execution plan, and a comparator $\mathcal{M}_c$ is to determine the efficiency of the plan.

\begin{algorithm}[!t]

\LinesNumbered

\eat{\linesnumbered}

\KwIn {Query set $Q$, hint set $H$, DBMS $D$, backbone LLM $\mathcal{M}_b$, larger LLM $\mathcal{M}_l$, lightweight model $\mathcal{M}_p$, comparator model $\mathcal{M}_c$, training epochs $E$.}

\KwOut {Updated model $\mathcal{M}_b$, $\mathcal{M}_p$, and $\mathcal{M}_c$.}

Initialize training set $S$;

\For{each query $q \in Q$ }{

    $q_c\leftarrow \mathcal{M}_l(q)$;

    \For {each $h \in H$}{

        Collect running time $t(p_q^{h})$ for $q$ with plan $p_q^h$ on $D$;

        Compose the absolute matching text $M$;

    }

    \For {each two hint $h_0, h_1 \in H$}{

        $S\leftarrow S \cup (q, q_c, h_0, h_1, p_q^{h0}, p_q^{h1}, M_0, M_1, l)$, where $l$ is the label determined from running time between $t(p_q^{h0})$ and $t(p_q^{h1})$;

    }

}

\While{$\#$ of epochs less than  $E$}{

    \For{each batch $B$ sampled from $S$}{

        \For{$(q, q_c, h_0, h_1, p_q^{h0}, p_q^{h1}, l)\in B$}{

            $e_q^{h0} \leftarrow \mathcal{M}_p(p_q^{h0})$; $e_q^{h1} \leftarrow \mathcal{M}_p(p_q^{h1})$;

            Apply $\mathcal{M}_b$ on $q_c$, $h_0$, $e_q^{h0}$ and $M_0$ to update $e_q^{h0}$;

            Apply $\mathcal{M}_b$ on $q_c$, $h_0$, $e_q^{h1}$ and $M_1$ to update $e_q^{h1}$;

            Apply $\mathcal{M}_c(e_q^{h0}, e_q^{h0})$ to predict the relative performance by Equation~\ref{eq:comparator_prob_train} and~\ref{eq:label_truth};

            Update parameters of model $\mathcal{M}_b$, $\mathcal{M}_p$, and $\mathcal{M}_c$ based on loss function by Equation~\ref{eq:loss1};
        }
    }
}

\caption{Training LLM4Hint \label{alg:training_LLM4Hint}}

\end{algorithm}

Algorithm~\ref{alg:training_LLM4Hint} consists of two major phases. The first phase focuses on data preparation. We translate each SQL query into the natural language using an LLM $\mathcal{M}_l$ like GPT-4. \eat{This translation is performed per query, minimizing API call fees. }We generate soft prompt for each plan using $\mathcal{M}_p$, and construct the matching prompt with explicit relationships. We collect the execution time for each generated plan and generate the labels by comparing the time consumption between pairs of plans.

In the second phase on training, we sample the training data from the training set $S$. Each instance includes two plans from the same query according to the construction rules. We then compose the prompt using the enhanced SQL text, the hint description, the soft prompt from the lightweight model, and the matching text, to produce the final plan embedding. By feeding two plan embeddings between pairs of plans into the comparator, we finally obtain the computed results, and optimize the parameters using the difference between these results and the ground-truth labels. Note that the backbone model $\mathcal{M}_b$ can either remain frozen or be fine-tuned using methods, like LoRA~\citep{lora}.

The inference process of LLM4Hint is described in Algorithm~\ref{alg:inference_LLM4Hint}, which handles a single query $q$ and can be easily extended to the whole query set. We first apply LLM $\mathcal{M}_l$ to generate the natural language representation of $q$. For each efficient hint, the corresponding hint-guided query execution plan is retrieved from the DBMS via {\it Explain} statement. The plan embedding is then generated by the backbone model based on the composed prompt in Line 6. For each pair of hints, the comparator predicts a performance label, which, for simplicity, is used as the score. Specifically, when the execution plan $p_q^{h0}$ generated by hint $h_0$ performs significantly better than the plan $p_q^{h1}$ from hint $h_1$, the score for $h_0$ is incremented by 3 (the label for $h_0$), and the score for $h_1$ is incremented by 1 (the label for $h_1$). The hint $h_s$ that achieves the highest cumulative score in the comparison matrix is selected as the best-performing hint. Finally, the selected hint $h_s$ is recommended to generate the final query execution plan.

\begin{algorithm}

\LinesNumbered

\KwIn {A query $q$, hint set $H$, DBMS $D$, backbone LLM $\mathcal{M}_b$, larger LLM $\mathcal{M}_l$, lightweight model $\mathcal{M}_p$, comparator model $\mathcal{M}_c$.}

\KwOut {Selected hint $h_s$ for $q$.}

$q_c\leftarrow \mathcal{M}_l(q)$;

\For {each $h \in H$}{

    Collect query execution plan $p_q^h$ on $D$;

    Compose the absolute matching relationship $m$;

}

$e_q^{h} \leftarrow \mathcal{M}_p(p_q^{h})$ for each $h\in H$;

Apply $\mathcal{M}_b$ on $q_c$, $h$, $e_q^{h}$ and $m$ to update $e_q^{h}$ for each $h\in H$;

Initialize score $S_h$ for each hint $h\in H$;

\For {each two hint $h_0, h_1 \in H$}{

    Apply $\mathcal{M}_c(e_q^{h0}, e_q^{h1})$ to predict the relative performance label $s$;

    $S_{h_0} \leftarrow s$, $S_{h_1} \leftarrow 4 - s$;

}

\KwRet{hint $h_s$ that maximizes the cumulative comparison score $S_{h_0}$.}

\caption{Inference using LLM4Hint\label{alg:inference_LLM4Hint}}

\end{algorithm}

The time complexity of the inference phase can be described as follows. Given the hint set $H$, we need to invoke a larger LLM $\mathcal{M}_l$ one time, run the backbone LLM $\mathcal{M}_b$ to generate the plan embeddings in $O(|H|)$ times, and invoke the comparator $\mathcal{M}_c$ on the pairs of plan embeddings in $O(|H|^2)$ times. As the computation of $\mathcal{M}_c$ can be done in parallel, the time cost of $\mathcal{M}_c$ can be reduced to $O(1)$ using the GPU acceleration. The overall time complexity is similar to the other pair-wise selection methods, like Lero~\citep{lero2023}.

\section{Experiment}
\label{sec-exp}

In this section, we first outline the details of the experimental setup used in our evaluation. Following that, we present the overall performance comparison of LLM4Hint to its competitors. We then illustrate the effectiveness of different components and conduct a deep investigation on the roles of LLM in LLM4Hint. Finally, we report the results for different backbone LLMs and training strategies.

\subsection{Experimental Setting}
\label{sec:setting}

\stitle{Dataset.} We utilize three widely used datasets to evaluate LLM4Hint: Join Order Benchmark (JOB)~\citep{goodplan2015}, TPC-DS~\citep{tpcds}, and Stack Overflow dataset~\citep{bao2021}. For these queries, existing methods always randomly sample queries to build the training and test set. We follow the same setting. As these queries actually come from the fixed number of templates, we also split the training and test sets across the templates to better evaluate the generalization of the method. We use different suffixes to distinguish these two splitting strategies, in which the suffix "-Q" performs splitting at the level of the queries, and the suffix "-T" performs splitting at the level of the templates.

\begin{itemize}

\item JOB dataset, derived from the IMDb database, comprises 33 query templates and a total of 113 queries. \textit{JOB-Q} splitting follows the approach~\citep{LOGER2023}, where one query from each template (1a$\sim$33a) is included in the test set, and the remaining queries are used for training. In \textit{JOB-T} splitting, we use the queries from 9 templates as the test set, and view the queries from the remaining 24 templates as the training set.

\item TPC-DS is a widely used benchmark for decision support. The data scale factor is set to 4 in our experiments. \textit{TPC-DS-Q} has its training and test set selected at the level of queries from a query subset, as the execution plans of certain TPC-DS queries contain operators that are not supported by some competitors. Specifically, TPC-DS-Q includes 20 templates selected from the total of 99 templates, with each template generating three distinct queries. Following a 3:2 ratio, we randomly divide the queries into training and test sets~\citep{LOGER2023}. As for \textit{TPC-DS-T}, we use all 99 templates, among which we allocate 80 queries for training and the rest for testing.

\item Stack Overflow dataset includes queries collected from real-world query workloads. Due to similarities among templates, we follow one template splitting strategy~\citep{LOGER2023}. Specifically, we select 10 queries from each of 8 templates to form the training set, and 5 queries from each of the remaining\eat{6} templates to construct the test set.

\eat{
 in generating the train/test set
}

\end{itemize}

\begin{table*}
    \centering
    \caption{Performance Comparison on JOB-Q, TPC-DS-Q and Stack.}
    \label{tab:allperformcompare}
    \begin{tabular}{ccccccc}
        \toprule
        \multirow{2}{*}{{Model}} & \multicolumn{2}{c}{{JOB-Q}} & \multicolumn{2}{c}{{TPC-DS-Q}} & \multicolumn{2}{c}{{Stack}}
        \\
        \cmidrule(lr){2-3} \cmidrule(lr){4-5} \cmidrule(lr){6-7}
        & {SU$\color{black}{\uparrow}$} & {GMRL$\color{black}{\downarrow}$} & {SU$\color{black}{\uparrow}$} & {GMRL$\color{black}{\downarrow}$} & {SU$\color{black}{\uparrow}$} & {GMRL$\color{black}{\downarrow}$} \\
        \midrule
        $O_0$ & 1.00 & 1.00 & 1.00 & 1.00 & 1.00& 1.00\\
        $O_1$ & 1.00 & 0.99 & 0.85 & 2.56 & $\approx$0 & -\\
        $O_2$ & 0.57 & 1.45 & 1.44& 0.84& 0.76 & 1.11\\
        $O_3$ & 0.63 & 4.17 & 0.88&2.47& $\approx$0 & -\\
        $O_4$ & 1.04 & 0.94 & 1.06&0.96& 1.16 & 0.90\\
        $O_5$& 0.23 & 7.21 & 0.09 &6.93& $\approx$0&-\\
        $O_6$& 0.87 & 1.24 & 0.29 &2.24& $\approx$0&-\\
        $O_7$& 0.23 & 7.26 & 0.09&6.45& $\approx$0&-\\
        $O_8$& 0.15 & 9.85 & 0.09 &5.41& $\approx$0&-\\
        $O_9$& 0.99 & 1.17 & 0.32 &2.04& $\approx$0& -\\
        $O_{10}$ & 0.64 & 4.13 &0.98 &2.26& $\approx$0&-\\
        $O_{11}$ & 0.59 & 1.42 & 1.60&0.79& 0.70 &1.15\\
        $O_{12}$ & 0.66 & 3.95 &0.98 &2.25& $\approx$0&-\\
        $O_{13}$ & 1.07 & 1.08 &0.33 &2.01& $\approx$0&- \\
        $O_{14}$ & 1.20 & 1.01 & 0.18&3.32& $\approx$0&- \\
        $O_{15}$ & 1.11 & 0.87 &0.51 &1.56& $\approx$0&- \\
        \hline
        $O_{h^{*}}$ &2.11&0.65&1.92&0.72&1.21&0.72\\
        \hline
        Bao & 1.32 & 0.88 & 1.75 & 0.81 & 1.08 & 0.91\\
        Lero  & 1.21 & 0.87 & 1.26 & 0.85 & 0.49 & 1.41\\
        FASTgres & 1.39 & 0.86 & - & - & 0.62 & 1.25\\
        \hline
        \eat{
        }
        LLM4Hint
        & \textbf{1.70}($\color{black}{\uparrow}$\textbf{22.0\%})
        & \textbf{0.79}($\color{black}{\downarrow}$\textbf{8.1\%})
        \eat{
        }
        & \textbf{1.86}($\color{black}{\uparrow}$\textbf{6.3\%})
        & \textbf{0.75}($\color{black}{\downarrow}$\textbf{7.4\%})
        & \textbf{1.16}($\color{black}{\uparrow}$\textbf{7.4\%})
        & \textbf{0.90}($\color{black}{\downarrow}$\textbf{1.1\%})
        \\
        \eat{
        }
        \bottomrule
    \end{tabular}
\end{table*}

\stitle{Competitors.} As LLM4Hint relies on the DBMS to produce candidate plans, we mainly compare LLM4Hint with other learned selection methods. In addition, we enumerate all hints, and locate the upper bound of the selection based method, which is denoted as the optimal optimizer.

\begin{itemize}

\item \textit{Optimizer with Different Hints ($O_i$)}: The query optimizer ($O_i$) represents the database's query optimizer with the $i$-th hint configuration selected from previous work~\citep{bao2021}. These hints focus on different combinations of join operators (\eg hash, merge, loop joins) and scan operators (\eg sequential, index, index only scans). Among these, $O_0$ denotes the database's default query optimizer.

\item \textit{Optimal Optimizer}: The Optimal Optimizer ($O_{h^{*}}$) is defined as the most efficient hint $h^{*}$ among all 16 hints for each query. It is a virtual ideal optimizer that represents the upper bound of the performance of all learned plan selection methods.

\item \textit{Bao}~\citep{bao2021}: Bao produces plans via hint combinations, and learns a value model to estimate the absolute execution time, from which the most efficient hint is selected.

\item \textit{Lero}~\citep{lero2023}: Lero takes the correction of estimated rows as the hint to produce various execution plans, and focuses on learning a model to compare the performance between two plans, due to the fact that the relative performance for plan selection is the key factor in the optimized plan generation.

\item \textit{FASTgres}~\citep{fastgres2023}: FASTgres clusters queries by table and join combinations, and uses traditional ML methods (\eg gradient boosting) to predict optimal hints based on filter predicates.

\end{itemize}

\stitle{Evaluation Metrics.} We choose two widely recognized metrics to assess the performance of the generated plan.

\begin{itemize}

\item \textit{SU (for Speedup)}: Given a query set $Q$, SU is the ratio between the time consumed by $Q$ using the default optimizer in the DBMS and the total time consumed using a specific optimizer. A higher SU score indicates a better optimizer.


\item \textit{GMRL (for Geometric Mean Relative Latency)}: Given a query set $Q$ and an optimizer $O$, GMRL is the geometric mean value of the ratios between the execution time of $O$ and that of the default optimizer for each query. A lower GMRL score indicates a better optimizer.

\end{itemize}

\stitle{Experiment Implementation.} We use PostgreSQL 13.10 as the DBMS running on an Ubuntu 22.04 server with Intel Xeon Gold 4210R CPU, 256GB memory, and one NVIDIA RTX A6000 GPU. We use the smallest GPT‑2 model (117M parameters) as the default backbone LLM in LLM4Hint. We select the top 16 efficient hint combinations, as identified by Bao~\citep{bao2021}. For the lightweight model, we mainly use the official source codes for QueryFormer~\citep{QueryFormer2022}. Both the lightweight model and the comparator are trained with the AdamW optimizer. The training procedure employs early stopping based on convergence criteria: training is halted either when the loss improvement falls below a fixed tolerance threshold (\eg 0.01) for a specified number of consecutive epochs (\eg 5), or after reaching a predefined maximum number of epochs (\eg 50), whichever occurs first. The optional larger LLM used in converting the SQL to the natural language is GPT-4~\citep{gpt4}. The source code \footnote{https://anonymous.4open.science/r/LLM4Hint-E60F/} of LLM4Hint is openly accessible.

\eat{When the parameters in GPT-2 are frozen, we train the rest of the models in 50 epochs. When the parameters of GPT-2 are fined-tuned using LoRA~\citep{lora}, we train the entire LLM4Hint in 30 epochs. }

\stitle{Experiment Design.} We first compare LLM4Hint with other competitors. We then conduct an ablation study to illustrate the fine-tuning cost and effects of different components, including the combined LLM-Lightweight model, the SQL rewriting to natural language, and the explicit matching text in the prompts. We further make a deep analysis on the role of LLM in the performance improvement of LLM4Hint. We finally conduct a study with various backbone models and training strategies to show that the improvement of LLM4Hint does not come from a specific LLM.

\eat{, as the main work of this paper is to introduce LLM into the query optimization task}

\subsection{Overall Performance Comparison between LLM4Hint and Competitors}
\label{sec:performance}

We first conduct the overall performance study of LLM4Hint and competitors on JOB-Q, TPC-DS-Q, and Stack in terms of SU and GMRL. It is important to note that in the Stack dataset, the optimizer with certain hints exceeds the statement timeout threshold, \ie 1,000,000 ms. Thus, we mark the SU values as "$\approx 0$" for the corresponding hints. Besides, due to lack of support for TPC-DS queries by the current version of FASTgres, we record the performance of FASTgres with a dash ("-") in the table.

We have the following observations from Table~\ref{tab:allperformcompare}: (i) The default DBMS optimizer $O_0$, whose SU and GMRL are both 1, is not always ideal for all cases. For example, if we choose the correct hint each time, the Optimal Optimizer $O_{h^{*}}$ can achieve an SU greater than 2 in JOB-Q, revealing the potential of learned selection methods. (ii) LLM4Hint demonstrates consistent improvements over all competitors in all datasets, which illustrates that LLM4Hint has certain generalization in handling different query sets. (iii) The performance gains of LLM4Hint vary across different datasets. The largest gain is observed in JOB-Q, where LLM4Hint achieves an SU of 1.70$\times$, indicating a 22.0\% improvement compared to the best competitor FASTgres, and a GMRL of 0.79, which is 8.1\% lower than the best competitor FASTgres.

\begin{table*}
    \centering
    \caption{Ablation Study for Major Components of LLM4Hint and Fine-Tuning Time (Time per Epoch in Minutes $\times$ $\#$ Epochs) }
    \label{tab:abl}
    \resizebox{\textwidth}{!}{
    \begin{tabular}{cccccccccccccccc}
        \toprule
        \multirow{2}{*}{{Model}} & \multicolumn{4}{c}{{Prompt Components}} & \multicolumn{3}{c}{{JOB-Q}}  & \multicolumn{3}{c}{{TPC-DS-Q}} &\multicolumn{3}{c}{{Stack}} \\
        \cmidrule(lr){2-5} \cmidrule(lr){6-8}  \cmidrule(lr){9-11} \cmidrule(lr){12-14}
        & Rewritten & SQL & Hint & Soft Prompt & {SU$\uparrow$}&{GMRL$\downarrow$}&{Time$\downarrow$} & {SU$\uparrow$}&{GMRL$\downarrow$}&{Time$\downarrow$} & {SU$\uparrow$}&{GMRL$\downarrow$}&{Time$\downarrow$}  \\
        \midrule
        \multirow{4}{*}{LLM4Hint}
        &&&  & \checkmark & 1.38 & 0.88 &10$\times$50 &1.33 & 1.22 &4$\times$24 & 1.00 & 0.93& 5$\times$18\\
        
        &\checkmark && \checkmark &  & 1.57 & 0.84 &  17$\times$50& 1.52 & 0.96 &5$\times$19& 1.01 & 0.95&16$\times$24\\
        
        & & \checkmark  & \checkmark & \checkmark & 1.54 & 0.87 &40$\times$34& 1.82 & 0.79 &15$\times$19& \textbf{1.16} & \textbf{0.90}&24$\times$12\\
        
        &\checkmark && \checkmark & \checkmark  & \textbf{1.70} & \textbf{0.79} &39$\times$41& \textbf{1.86} & \textbf{0.75} &13$\times$27& \textbf{1.16} & \textbf{0.90} &42$\times$12\\
        \bottomrule
    \end{tabular}
    }
\label{tab:ablation}
\end{table*}

\subsection{Effectiveness of Different Components in LLM4Hint and Fine-Tuning Cost}

We study the fine-tuning time cost and effects of major components in LLM4Hint, including the combined LLM-Lightweight model, SQL rewriting using a larger LLM, and explicit matching text in the prompts. For the first two tests, we use query-level splitting, while for the third test, we use the template-level splitting strategy, as the explicit matching text plays a positive role in the generalization.

\stitle{The Role of Combined LLM-Lightweight Model.} Table~\ref{tab:abl} presents the results of LLM4Hint in the ablation experiments by removing certain parts from the prompt, among which "Rewritten" means the rewritten SQL text using a larger LLM, "SQL" means the original form of the SQL statement, and "Soft Prompt" means the soft prompt for the execution plan. Usually, we select only one form from "Rewritten" and "SQL". 

We mainly observe the roles of the combined LLM-Lightweight model from the first two rows. The first row reports the results of using only the lightweight model to produce the plan embedding. We can see that the results have a significant performance gap compared to the best one. As mentioned previously, while the lightweight model strives to capture the underlying distribution, it lacks the capacity to model the selectivity involving complex query features, including the "\textit{LIKE}" and "\textit{IN}" clauses. Additionally, the lightweight model achieves the shortest training time, due to its limited number of parameters and the short input token sequences of the LLM. 

The second row indicates the results using the LLM model only. We can observe that the LLM provides better results than the lightweight model. For example, the LLM can achieve an SU of 1.57 and 1.52 on the JOB-Q and TPC-DS-Q, respectively, both exceeding optimizers with nearly all hints ($O_0$ to $O_{15}$). However, using only textual input lacks certain internal database knowledge (such as data distribution), and such a query-driven method only learns the relationships between query expression and optimization hints during training. We also notice that the LLM only has little to no effect in the Stack workloads, which illustrates the need to combine the SQL text features and the soft prompt from the lightweight model.


\stitle{The Role of Rewritten SQL.} Rewritten SQL significantly boosts LLM4Hint's performance, especially on JOB-Q and TPC-DS-Q datasets. Using the smallest GPT-2 (117M parameters), the model struggles to understand raw SQL text and may fail to identify key components like tables. In contrast, when provided with SQL rewritten by a larger model like GPT-4, GPT-2 can recognize tables and predicates accurately. This highlights the importance of rewritten SQL for small LLMs, which offer low fine-tuning costs and no access fees in query optimization. This improvement is supported by the results in Table~\ref{tab:abl}. By comparing the third and fourth rows, we observe that introducing the rewritten SQL component yields a significant performance gain across most benchmarks—most notably on JOB-Q (from 1.54 to 1.70 SU, 0.87 to 0.79 GMRL) and TPC-DS-Q (from 1.82 to 1.86 SU, 0.79 to 0.75 GMRL)—while only incurring a marginal increase in fine-tuning time.


Besides, the rewritten SQL text has little or no effect on the Stack test workloads, as shown in Table \ref{tab:abl}. We guess that it is likely because all Stack queries share similar query patterns. Thus, either the raw SQL statement or the rewritten SQL statement only produces a similar text token sequence for the LLM. In such cases, the database statistics and the cardinality of the query operators become the major factors in differentiating different execution plans, whose roles are also observed from the first two rows of the table.

Regarding the fine-tuning cost, the SQL rewritten in natural language incurs longer time as shown in Table~\ref{tab:abl}. This is because the natural language representation typically contains more tokens than the original SQL, resulting in an expanded context for the underlying LLM and consequently increased fine-tuning cost. However, we believe that such an additional cost is justified due to the significant performance improvement in execution plans in most cases.

\eat{Besides, the rewritten SQL text has little to no effect on the Stack, as seen in Table \ref{tab:abl}. This is likely because most optimizers with hints in the Stack dataset perform abysmal, with acceleration rates close to zero. In such cases, the model's primary task shifts to distinguishing between hints and query plans, making a deeper understanding of SQL context less essential. Thus, the benefits of rewritten SQL become marginal in this scenario.}

\begin{table}[b]
    \centering
    \caption{Ablation Study of Matching Relationships in LLM4Hint}
    \begin{tabular}{cccccccc}
        \toprule
        \multirow{2}{*}{{Model}} & \multicolumn{2}{c}{{Matching}} & \multicolumn{2}{c}{{JOB-T}} & \multicolumn{2}{c}{{TPC-DS-T}} \\
        \cmidrule(lr){2-3} \cmidrule(lr){4-5} \cmidrule(lr){6-7}
        & REL & ABS & {SU$\uparrow$}&{GMRL$\downarrow$} & {SU$\uparrow$}&{GMRL$\downarrow$} \\
        \midrule
        \multirow{3}{*}{LLM4Hint}
        &&& 1.25 & 1.06 & 2.86 & 1.09 \\
        &\checkmark& & 1.69 & 0.89 & 2.97 & 0.90 \\
        && \checkmark & \textbf{1.79} & \textbf{0.88} & \textbf{3.00} & \textbf{0.89} \\
        \bottomrule
    \end{tabular}
\label{tab:match}
\end{table}

\stitle{The Role of Explicit Matching Text.} The explicit matching text, which links the table embedding from the lightweight model and the corresponding table in the SQL query, can not only facilitate model training, but also enhance generalization ability. For example, when table aliases are changed in the SQL expression, the model can still make a proper decision, as these different table aliases are related with the same table embeddings. To assess the generalization ability of LLM4Hint, we follow the splitting strategy based on the templates, in which the test queries do not belong to any templates of the training queries. Furthermore, we conduct validation tests to evaluate the impact of both absolute (ABS) and relative (REL) relationships in the explicit matching text. These tests incorporate all remaining prompt components, including the rewritten query text and the soft prompt generated by the lightweight model.

From the results in Table~\ref{tab:match}, we observe that the matching text can significantly enhance the performance of LLM4Hint. For example, integrating the absolute matching relationship enables LLM4Hint to achieve an SU of 1.79 on JOB-T in the 3-rd row, marking a 43.2\% improvement compared to the variant without matching. We also notice that the baseline method without the matching relationship module in the first row has achieved good performance in TPC-DS-T with an SU of 2.86, which is even larger than that on TPC-DS-Q. This may be partially due to the presence of certain long queries in the TPC-DS-T test set that offer larger SU gains. We can see that GMRL is bigger than 1 in this case. By combining these results, we can see the role of the matching relationships module in enhancing the generalization ability of LLM4Hint, \ie to demonstrate the capacity to adapt to unseen query templates.

Additionally, we find that the absolute matching strategy achieves better results than the relative relationship strategy on both query sets. The purpose of the introduction of matching text is to link the table name in the SQL text to the corresponding soft prompt. Compared to relative matching, absolute matching can reach the goal in an easier way, as the similarity computation in the transformer can discover such a relationship using the same table name and the same embedding. The index used in the relative relationship, although can be exploited by larger LLMs, poses difficulty for moderate-sized GPT-2 in understanding such a cross relationship.

\subsection{Deep Analysis on the Role of LLM}

We further perform a deep analysis on the role of LLM in performance improvement, including the type of queries that benefit from the LLM, and the way in which the LLM improves the effectiveness of the plan selection. After a deep investigation between LLM4Hint and the variant of LLM4Hint without LLM in the JOB-Q data set, we have the following two findings:

\begin{figure*}[!t]
  \centering
  \includegraphics[width=\linewidth]{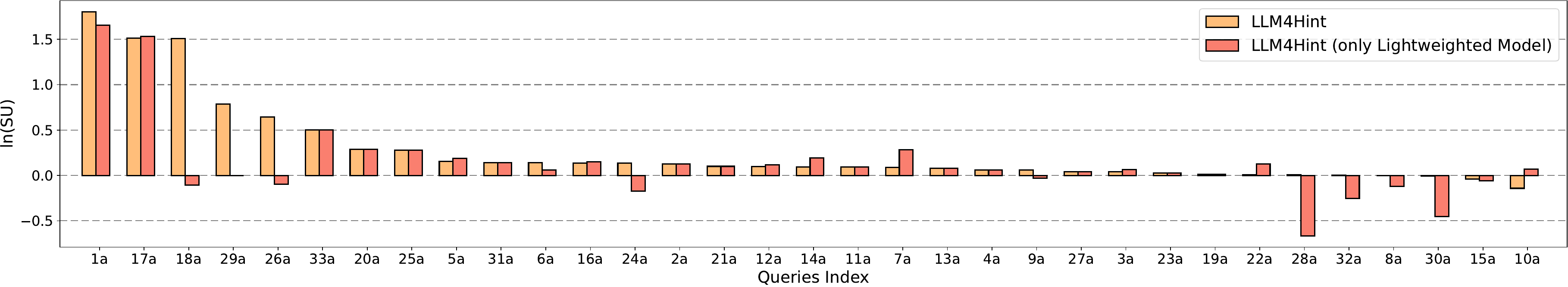}
  \caption{Speedup on JOB-Q Test Workload for Each Query}
  \label{fig:like}
\end{figure*}

\stitle{The SQL with more complex patterns, especially with string-related predicates, can gain more benefits from LLM.} We report the SU for each test query of LLM4Hint and its variant with the lightweight model only. The $Y$-axis in Figure~\ref{fig:like} takes the log-operation on SU to show which query is optimized, \ie positive score means that the optimized plan outperforms the plan from the default optimizer. The results show that: (i) most of test queries have been successfully optimized using LLM4Hint; (ii) LLM4Hint exhibits more substantial improvements in SU on queries 18a, 24a, 26a, 28a, 29a, 30a, and 32a compared to the lightweight model. LLM4Hint also shows notable improvements on queries 6a, 8a and 9a.

We further examine queries with significant performance improvements, and find that queries 29a, 28a, and 30a are the three longest in the test set, among which queries 28a and 30a optimized by the lightweight model cannot have positive optimization effects. In fact, these queries face high challenges due to multiple occurrences of complex predicates, such as "\textit{LIKE}" and "\textit{IN}", with 4 to 5 occurrences in each query. Besides, the "\textit{LIKE}" and "\textit{IN}" predicates have multiple constant strings as condition in their expression: 28a with 24 strings, 30a with 20 strings, and 29a with 16 strings. These complex patterns and large amounts of textual information make it difficult for lightweight models to optimize effectively~\citep{QueryFormer2022}.


We believe that the text understanding of LLM becomes the key complementary to the lightweight model in capturing the relationship between complex query patterns and optimization hints. For example, in queries involving several "\textit{LIKE}" conditions, LLM4Hint might infer the implicit relationships between the complex query predicates and the choice of indexes or filtering methods in the execution plan. Similarly, for queries with multiple "\textit{IN}" clauses, the model may account for the size and complexity of the value sets being processed, leading to more accurate cost predictions and plan selections. These findings suggest that LLM4Hint excels at handling queries with rich semantic predicates and complex structural patterns.

\eat{We believe that the text understanding of LLM used in the LLM4Hint, which becomes the key its complementary to the Lightweight model in capture the distribution of the string-related query predicate. For example, in queries involving several LIKE conditions, LLM4Hint can infer that the matching pattern might significantly impact the choice of indexes or filtering methods in the execution plan. Similarly, for queries with multiple IN clauses, the model can account for the size and complexity of the value sets being processed, leading to more accurate cost predictions and plan selections. These findings suggest that LLM4Hint excels at handling queries that are semantically rich and structurally complex. Utilizing SQL text enhances LLM4Hint's understanding of query behavior, allowing LLM4Hint to make more informed optimization decisions. This capability is particularly beneficial for queries where traditional optimizers or lightweight models may struggle due to the inability to model textual nuances.
}

\stitle{LLM4Hint simplifies the comparator with the easily-distinguished embeddings.} To further investigate how LLM works in LLM4Hint, we begin by examining the differences in the plan embeddings generated by LLM4Hint and the lightweight model on JOB-Q, before feeding these embeddings into the comparator. Obviously, the comparator can produce better results if the embeddings for efficient plans can be distinguished from those for poor plans.

\begin{figure}[t]
    \centering
    \subfloat[Query IDs of Embeddings]{\includegraphics[width=0.25\textwidth]{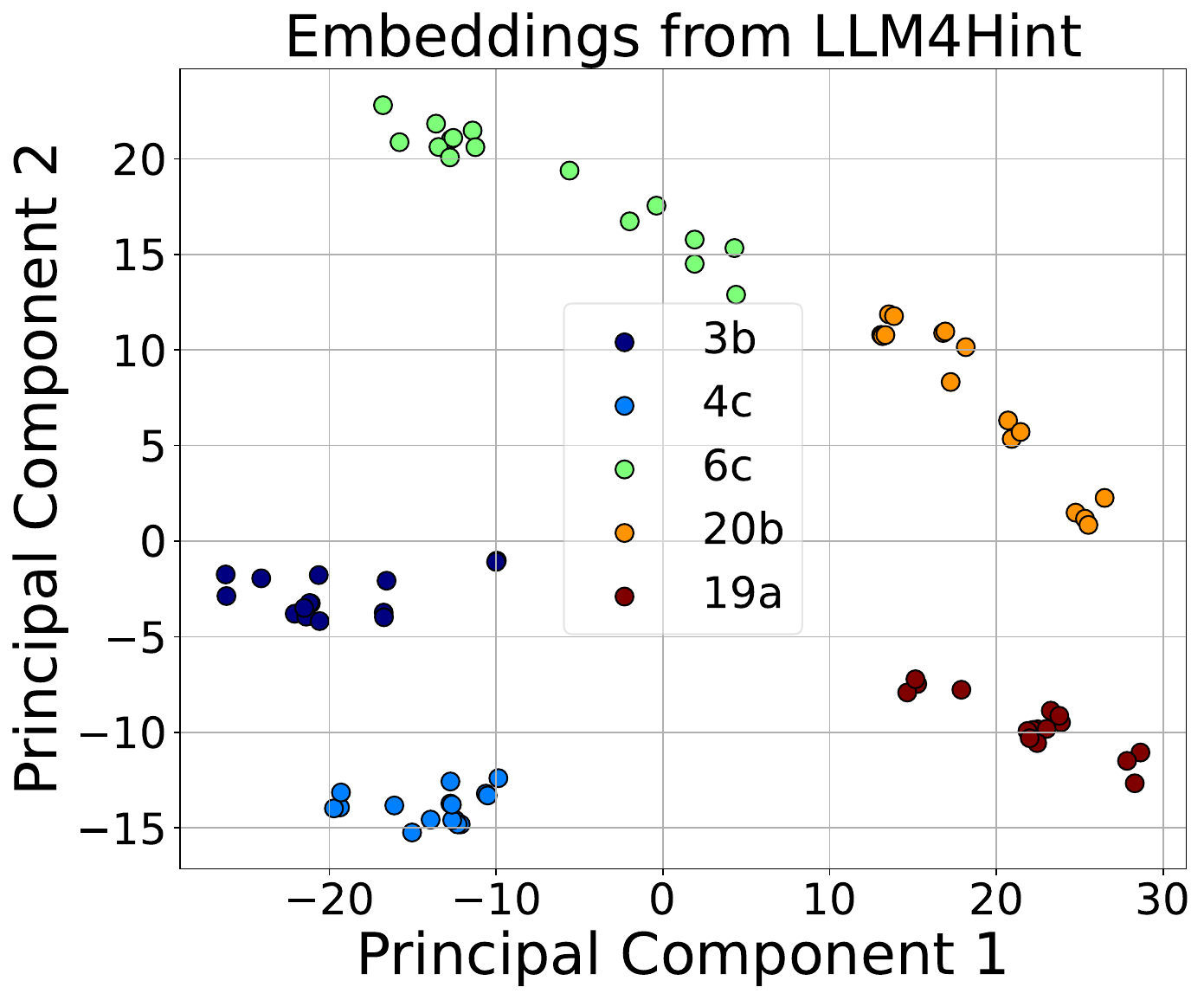}\label{fig:emb_q}}
    \subfloat[Performance Labels of Embeddings]{\includegraphics[width=0.25\textwidth]{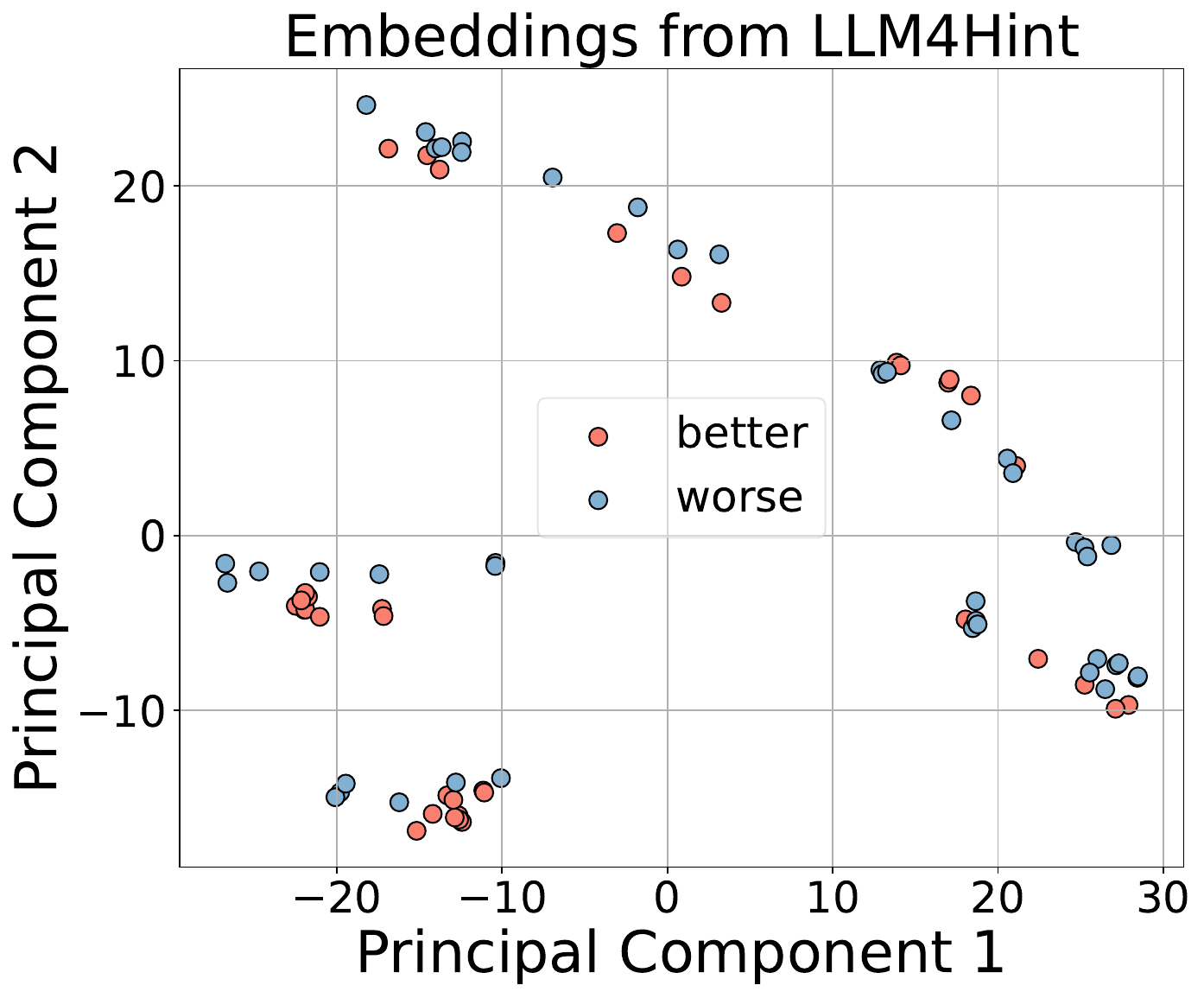} \label{fig:emb_best}}
    \\
    \subfloat[Query IDs of Embeddings]{\includegraphics[width=0.25\textwidth]{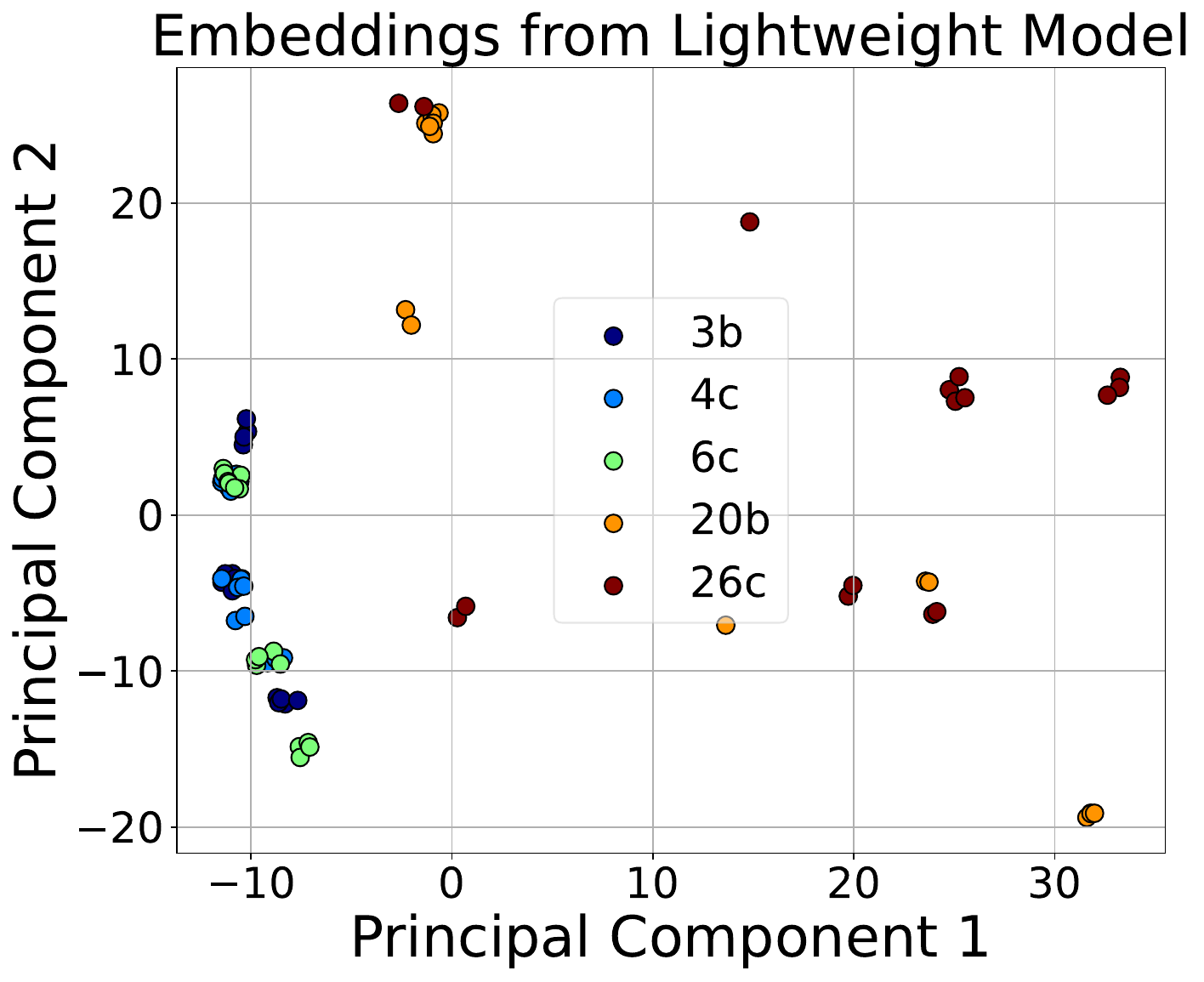}\label{fig:emb_qf_q}}
    \subfloat[Performance Labels of Embeddings]{\includegraphics[width=0.25\textwidth]{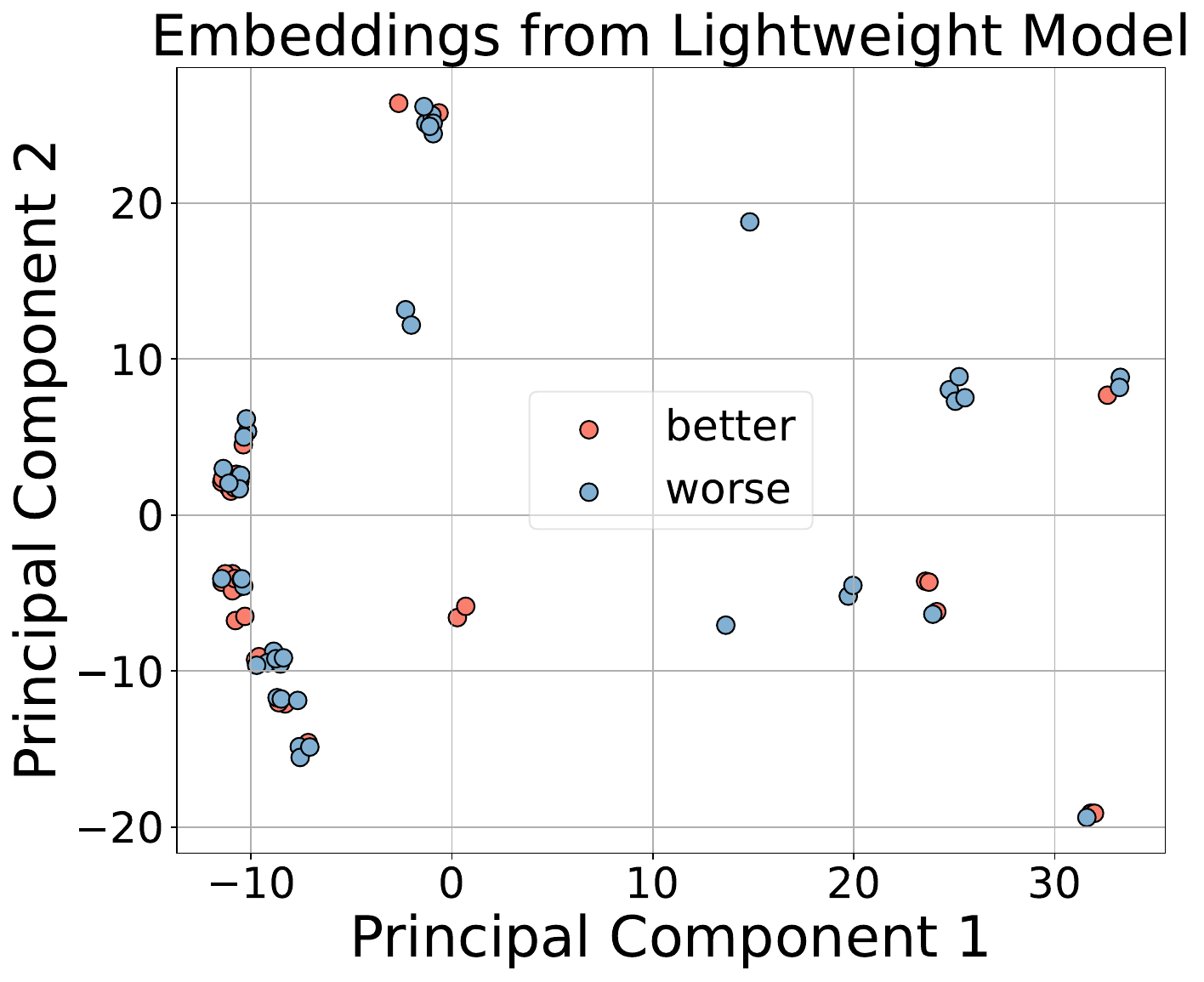}\label{fig:emb_qf_best}}
    \caption{Visualization of Embeddings by LLM4Hint or the Lightweight Model Only}
    \label{fig:emb}
\end{figure}

We perform a visualization study on the generated plan embeddings as follows. We randomly choose 5 queries, guide the generation of their execution plans according to all 16 different hints, and obtain the plan embeddings $e_c$ using the fully trained LLM4Hint. Similarly, we obtain the plan embeddings $e_q$ using the lightweight model only. Finally, we employ Principal Component Analysis (PCA) to project $e_c$ and $e_q$ into separate two-dimensional spaces. Each node represents an execution plan with two attributes. In addition to the query ID, each plan is annotated with its performance labels. Plans with execution times close to the optimal are labeled as "better", and the others are labeled as "worse".

We present the visualized results in Figure~\ref{fig:emb}. The upper part illustrates the embeddings from LLM4Hint, where the left sub-figure is from the viewpoint of query IDs, and the right sub-figure is from the viewpoint of the performance labels. The lower part illustrates the embeddings from the lightweight model only. By combining the left and right sub-figures, we can observe that compared to the embeddings produced by the lightweight model, the embeddings obtained using LLM4Hint exhibit stronger clustering with respect to each individual query, as the results inside five distinct categories with relatively clear boundaries. This clarity in distinction makes it easier for the following comparator to identify the more efficient plans, which then results in better results in the final hint recommendation.

\subsection{Varying Backbone LLMs and Training Strategies}
\label{sec:llm_b}

Finally, we choose different LLMs as the backbone model and explore training strategies to alleviate concerns that the performance improvement of LLM4Hint comes from a specific LLM. Besides the default GPT-2 mentioned before, we test two other representative LLMs on the JOB-Q dataset, among which \textbf{Llama 2-Chat}~\citep{llama2} contains 7 billion parameters designed for language generation, and \textbf{Qwen2-VL}~\citep{wang2024qwen2} also contains 7 billion parameters primarily for multi-modal tasks. We choose this model as there is also a need to establish cross-modal alignment between the SQL text and the embedding from lightweight model in LLM4Hint.

\stitle{Comparison of Parameters Frozen and Model Fine-tuning.} We attempt to build the learned query optimizers within limited computing resources, \ie using one single GPU card. Considering the given resource, we can fine-tune with LoRA~\citep{lora} or keep parameters unchanged on the moderate-sized GPT-2, while only freezing the parameters in the other two relative larger LLMs.

The results in the first two rows in Table \ref{tab:llm_big} show that the fine-tuned GPT-2 significantly outperforms the GPT-2 with frozen parameters. During the fine-tuning using LoRA, a small number of extra parameters are introduced to enhance the capability to understand SQL statements. In addition, the cross-relationship between the LLM and the lightweight model can be easily established, as parameters in both models can be adjusted simultaneously.

\stitle{Comparison of Different Backbone Models.} We can see that the other two LLMs also achieve performance improvement in the optimized plans, verifying that the specific LLM is not the key factor to the improvement. Due to large parameters, the other two LLMs have their parameters frozen in the training of LLM4Hint.

\begin{table}[t]
    \centering
    \caption{Performance Varying Training Strategies and Backbone LLMs on JOB-Q and Fine-Tuning Time (Time per Epoch in Minutes $\times$ $\#$ Epochs)}
    \begin{tabular}{ccccccccc}
        \toprule
        \multirow{2}{*}{{\makecell{Backbone of\\LLM4Hint}}} & \multicolumn{2}{c}{{Training Strategy}} & \multicolumn{3}{c}{{JOB-Q}} \\
        \cmidrule(lr){2-3} \cmidrule(lr){4-6}
         & Freeze & Fine-tune & {SU$\uparrow$}&{GMRL$\downarrow$} & Time$\downarrow$\\
        \midrule
        \multirow{2}{*}{GPT-2}
        & & \checkmark & 1.70 & 0.79&39$\times$41\\
        & \checkmark & & 1.29 & 0.89&36$\times$31\\
        \midrule
        Llama 2-Chat& \checkmark & & \textbf{1.77} & \textbf{0.75}&  632$\times$7\\
         Qwen2-VL& \checkmark & & 1.57 & 0.83&644$\times$10\\
        \bottomrule
    \end{tabular}
\label{tab:llm_big}
\end{table}

For the Llama 2-Chat model, we can see that it achieves the best results, with 37\% improvement in terms of SU compared to GPT-2 with parameters frozen, and even a slight improvement in terms of SU compared to fine-tuned GPT-2. We believe that such an improvement is mainly due to the large number of parameters in Llama 2-Chat, which are nearly 60 times larger than those in GPT-2.

Qwen2-VL outperforms GPT-2 but falls short of Llama 2-Chat. The knowledge encoded in Qwen2-VL covers text, image, and their relationships, providing a weak understanding of the SQL statement. Additionally, although Qwen2-VL has knowledge of cross-modal relationships, this knowledge differs significantly from the relationships between the LLM and the lightweight model, making knowledge transfer challenging based on current results.


We also observe significant differences in fine-tuning time. For instance, training LLM4Hint with Llama 2-Chat or Qwen2-VL takes at least twice the time on GPT-2. Such a discrepancy underscores that practical implementation requires careful consideration of both model performance and fine-tuning costs when selecting backbone LLMs.

\eat{ when using a single NVIDIA A6000 GPU}

\section{Conclusion}
\label{sec-conclusion}

In this paper, we introduce LLM4Hint, a pioneering framework that leverages the LLM for query hint recommendation in offline DBMS query optimization. To overcome the mismatch between the LLM and the query optimization task, LLM4Hint adopts a combined LLM-Lightweight model, in which the LLM is used to comprehend complex SQL queries, and the lightweight model is used to capture the tree-structured execution plan with rich-numeral features. We also propose several optimization strategies, including the SQL rewriting into natural language by a commercial LLM to aid the moderate-sized LLM in understanding the complex SQL queries, and the explicit matching prompt to lower the costs in alignment between the two models and enhance generalization ability. Extensive experimental studies across various datasets demonstrate that LLM4Hint improves the efficiency and generalization of the optimizer compared to other state-of-the-art learned selection methods.


\end{document}